\documentclass[sigconf]{acmart}
\usepackage{amsmath,bm}
\usepackage{adjustbox}
\usepackage{multicol}

\usepackage[utf8]{inputenc} 
\usepackage[T1]{fontenc}    
\usepackage{hyperref}       
\usepackage{url}            
\usepackage{booktabs}       
\usepackage{amsfonts}       
\usepackage{nicefrac}       
\usepackage{microtype}      
\usepackage{subfig}
\AtBeginDocument{%
  }

\copyrightyear{2025}
\acmYear{2025}
\setcopyright{acmlicensed}
\acmConference[WWW Companion '25]{Companion
Proceedings of the ACM Web Conference 2025}{April 28-May 2, 2025}{Sydney,
NSW, Australia}
\acmBooktitle{Companion Proceedings of the ACM Web Conference 2025 (WWW
Companion '25), April 28-May 2, 2025, Sydney, NSW, Australia}
\acmDOI{10.1145/3701716.3717567}
\acmISBN{979-8-4007-1331-6/2025/04}

\begin{document}
\title{Multimodal semantic retrieval for product search}

\author{Dong Liu}
\authornote{\textbf{Both authors contributed equally to this research.}}
\email{liuadong@amazon.com}
\affiliation{%
  \institution{Amazon}
  \country{Luxembourg}
}
\author{Esther Lopez Ramos}
\email{rsther@amazon.es}
\authornotemark[1]
\affiliation{%
  \institution{Amazon}
  \country{Spain}
}

\renewcommand{\shortauthors}{Dong et al.}

\begin{abstract}
Semantic retrieval (also known as dense retrieval) based on textual data has been extensively studied for both web search and product search application fields, where the relevance of a query and a potential target document is computed by their dense vector representation comparison. Product image is crucial for e-commerce search interactions and is a key factor for customers at product explorations. However, its impact on semantic retrieval has not been well studied yet. In this research, we build a multimodal representation for product items in e-commerce search in contrast to pure-text representation of products, and investigate the impact of such representations. The models are developed and evaluated on e-commerce datasets. We demonstrate that a multimodal representation scheme for a product can show improvement either on purchase recall or relevance accuracy in semantic retrieval. Additionally, we provide numerical analysis for exclusive matches retrieved by a multimodal semantic retrieval model versus a text-only semantic retrieval model, to demonstrate the validation of multimodal solutions. 
\end{abstract}

\maketitle

\section{Introduction}
Product search (or e-commerce search) is primarily based on lexical retrieval solution consisting of matching and ranking phases \cite{schutze2008introduction}, where product match is conducted by keyword term matching between a search query and a product document (or text input). Lexical solution relies on the overlapping keywords between the customer query and the product terms. However it does not capture customer preferences in context of product search and also cannot model the semantic relevance when a search query does not contain overlapping keywords with potential relevant products. 

The limitation of lexical retrieval is relieved by semantic retrieval solutions. Starting from deep semantic search model \cite{huang2013learning}, the two-tower architecture became a promising solution in semantic retrieval track. 
The recent advances in language models such as BERT unblock the community to rely on learning efficient representation in a latent vector space \cite{kenton2019bert}.
After introducing BERT into the semantic representation for a query or document, bi-encoder structured model got more popular in the domain \cite{zhan2020repbert,nie2020dc, khattab2021relevance}. Instead of relying on the keywords overlapping, a query or document is represented by an embedding vector. When it comes to the relevance, a relevance score is directly computed by comparing their the embedding representations. This approach is also referred as dense retrieval, since query and product are represented by dense vectors (embeddings). It has been popularly used in product search \cite{lakshman2021embracing, li2021embedding}. 

For e-commerce search, the product title is typically used for building the representation \cite{lakshman2021embracing, chang2021extreme}. Although product titles are a typical information source for customers as they explore products in online shopping, product images are key information source for them in judging if a product is relevant. Product image is a key factor in e-commerce \cite{christof2020picture}. 

The advent of multimodal models, such as ViLT \cite{kim2021vilt}, BLIP \cite{pmlr-v162-li22n}, Flamingo \cite{alayrac2022flamingo}, and Uni-Perceiver \cite{zhu2022uni}, has demonstrated the potential of jointly leveraging textual and visual modalities for various retrieval tasks. ViLT integrates text and vision using transformer architectures without heavy reliance on pre-computed visual embeddings, achieving efficient fusion. BLIP extends multimodal reasoning through bootstrapped pretraining, significantly advancing multimodal understanding and retrieval. Flamingo, with its few-shot learning capabilities, excels in tasks requiring multimodal alignment, while Uni-Perceiver unifies multimodal processing in a single transformer architecture, simplifying training across modalities.

Despite the importance of product images in e-commerce, image or visual based retrieval solutions have not been well explored yet. Existing models either prioritize general-purpose retrieval or focus on scenarios with abundant paired multimodal datasets, such as image captions. Although there is early work on using the hashing for image retrieval \cite{lin2021deep, zhang2021binary}, our target is to use it for relevant product retrieval. 
Besides, both textual information such as product title and product images are both important sources and discarding one over the other is information loss.

Our work fills this gap by introducing multimodal semantic retrieval models tailored for product search, addressing unique challenges like misalignment between independently trained text and visual encoders and the computational efficiency required for large-scale product catalogs. Closely related to our work,
\cite{zhu2024bringing} brings multimodality to e-commerce. In this approach they train both a 3-tower model where the query is an image (one visual encoder to encode query image, one visual encoder tower to encode product image and one text encoder to encode product text) and a 4-tower model where the query itself is multimodal, (i.e. the query consists of text and image), showing better image to image matching problem. Additionally, recent work \cite{zhu2024bringing} proposed a 3-tower model and compared between RoBERTa and a product BERT model together with visual encoder by using one additional encoder to fuse text and image embeddings. It shows improved relevance accuracy compared to publicly pretrained RoBERTa encoder based on MaxSim score function. However, MaxSim score is known as not being scalable with large index size and is not supported by dominant approximate k-nearest neighbour indexers. 
Dot product or cosine similarity (normalized dot product) is a simpler score function compared to MaxSim, and is computation-efficient in retrieval with large scale index. How multimodal representation makes a difference in product retrieval under scalable relevant scoring and indexing is not clear yet. Additionally, it is not clear that product retrieval task gets impacted by mis-alignment issue between separately trained text encoder and visual encoder.  

Our work has the following contributions:
\begin{itemize}
    \item We analyze different strategies to integrate visual transformers to text-only semantic retrieval model, and study multiple models to understand how product image makes a difference to product retrieval task. We study text-only, 4-tower multimodal semantic model and 3-tower multimodal semantic model for product retrieval task, helping to understand language-visual alignment for product search problem.
    \item Integrating a public pretrained model can contribute to the relevance scores significantly in 4-tower model setting.  We show how performance of a fine-tuned 3-tower semantic models can get close to that of a 4-tower model with some degradation. 
    \item Our developed solutions allows large-scale retrieval efficiently. Our benchmarks are under index size (over millions of product items) with straightforward Cosine similarity score function. The effectiveness of our methods are measured under purchase recall and detailed breakdown of relevance accuracy metrics. Additionally we carry out analysis on how good the exclusive matched products are for our developed multimodal models.  
\end{itemize}




\section{Preliminary}

To provide the context for the multimodal study of this work, we start with preliminary of semantic retrieval problem itself, followed by the description of the baseline models. Note we would be interleaving the usage of query and search query to denote text keywords in a search session. 

\subsection{Semantic Retrieval Problem}
Dual-encoder architecture (also known as bi-encoder network or 2-tower model) is a widely used model for semantic retrieval task. The work solution was initially proposed for web-search as the deep structure semantic models (DSSM) \cite{huang2013learning} where a multiple layer perception (MLP) model was used for encoding text input instead of BERT model. Such model can be trained by the typical normalized
temperature-scaled cross entropy loss (NT-Xent) which was initially proposed for visual representation learning in \cite{chen2020simple}. In the context of semantic retrieval, the NT-Xent loss can be coined as 
\begin{equation}
\label{eq_infonce}
    L = - \log \frac{\exp(s(\bm{q}_i, \bm{d}^{+}_i))}{\sum_j \exp(s(\bm{q}_i, \bm{d}^{-}_j))}
\end{equation}
where $\bm{q} = \phi_{q}(\bm{x}_{query})$ is the query encoder model $\phi_q$ mapping the query in text sequence $\bm{x}_{query}$ into vector space $\bm{q} \in \mathbb{R}^{N}$. Here $N$ is the dimension of the semantic vector space. $\bm{q}$, also named the embedding of query, is the semantic vector representation of search query. Similarly, $\bm{d} = \phi_{d}(\bm{x}_{doc})$ with  $\bm{d} \in \mathbb{R}^{N}$, is the semantic vector representation of a potential target document (or product text in product search context) via the document encoder $\phi_d$. We use subscript to index the embedding vector, i.e. $\bm{q}_i$ denotes the $i$-th input query embedding representation and $\bm{d}_j$ the $j$-th document embedding representation. $\bm{d}_j^{+}$ denotes the $i$-th relevant target representation and $\bm{d}_j^{-}$ is $j$-th irrelevant target representation for a given query.

\subsection{Baseline Models}

We introduce the baseline models in this section and start with the language-image model. Open AI CLIP (Contrastive Language–Image Pre-training) model \cite{radford2021learning} was initially coined for general visual concept problems and showed superior “zero-shot” capability. Specifically, CLIP is built upon a vision transformer as image encoder and a transformer as text encoder \cite{dosovitskiy2020image}, respectively. The model is trained on image-text pairs data to align the text and image representations. For this work, we adapted a pretrained CLIP model \cite{fang2023data} with 5 billion images, which is not fine-tuned on e-commerce dataset.

As for text-only baseline model, we adapt a BiBERT model, i.e. a bi-encoder model with two towers (one encoder tower for query text and one for document text encoding). The BiBERT model is
trained on query-product pairs associated with purchase-interest patterns. In training phase, the loss function Eq.\ref{eq_infonce} is applied.
Each of BiBERT model contains 2-transformers layer and 4 attention heads. Most of the query and product document encoder towers weights are shared except for the LayerNorm and Projection Layer. 

\section{Fusion Models}

In this section, we introduce the multi-modality models that we investigate in this work. The text encoder tower and visual encoder tower of CLIP model are aligned during training in the semantic vector space. But it may not be true when comparing a text tower of BiBERT and the visual tower of CLIP, since both models were trained independently. Therefore, we studied two types of models, i.e. 4-tower Multimodality Model (4tMM) and 3-tower Multi-modality Model (3tMM). The comparison of the two architecture types may help us understand if the semantic vector space misalignment can be migrated by further model fine-tuning.
\begin{figure*}
  \centering
  \centering
    \subfloat[\centering 4-tower based Multimodal Model ]{{\includegraphics[width=5cm]{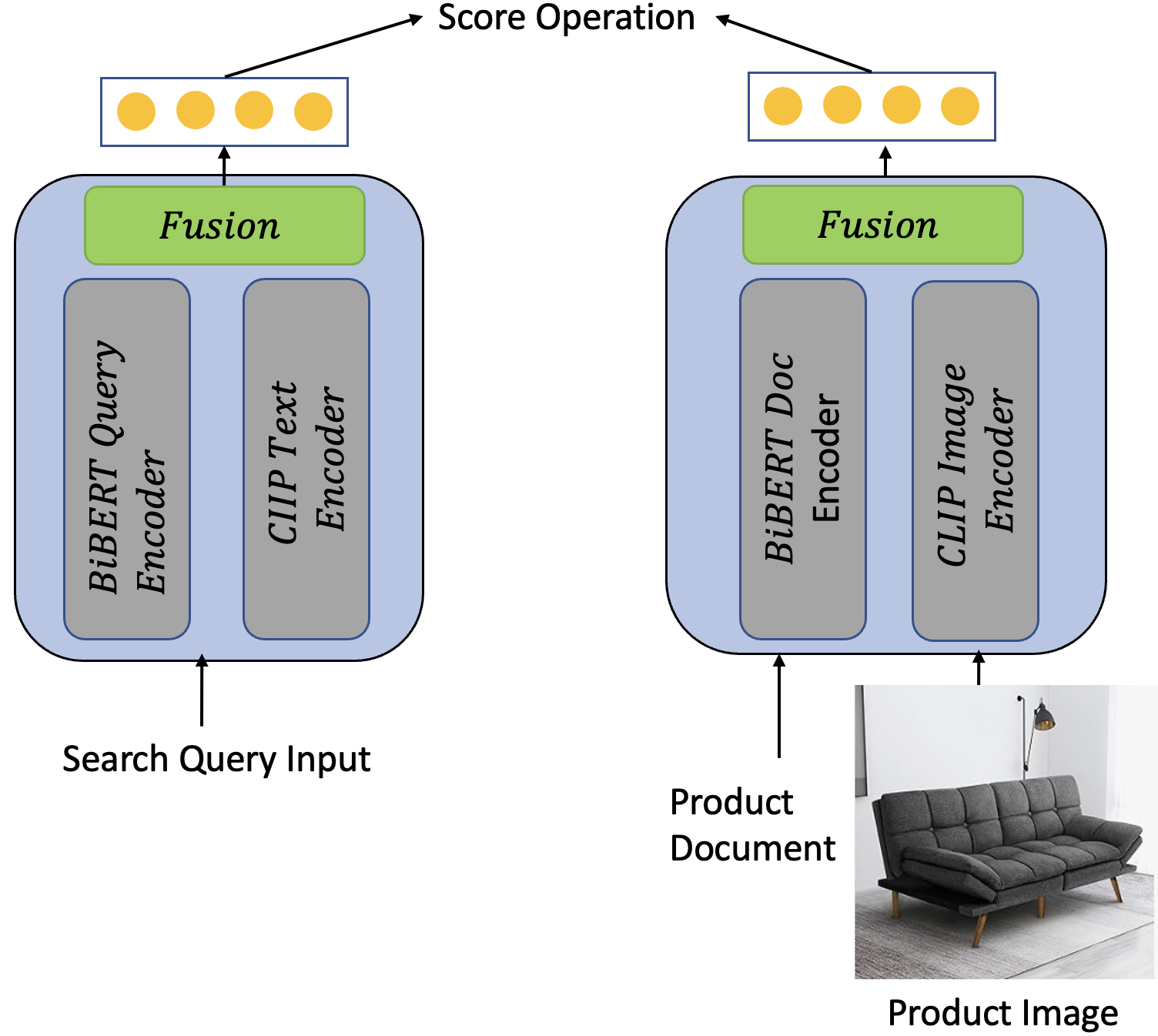} }}%
    \qquad
    \subfloat[\centering 3-tower based Multimodal Model ]{{\includegraphics[width=4.6cm]{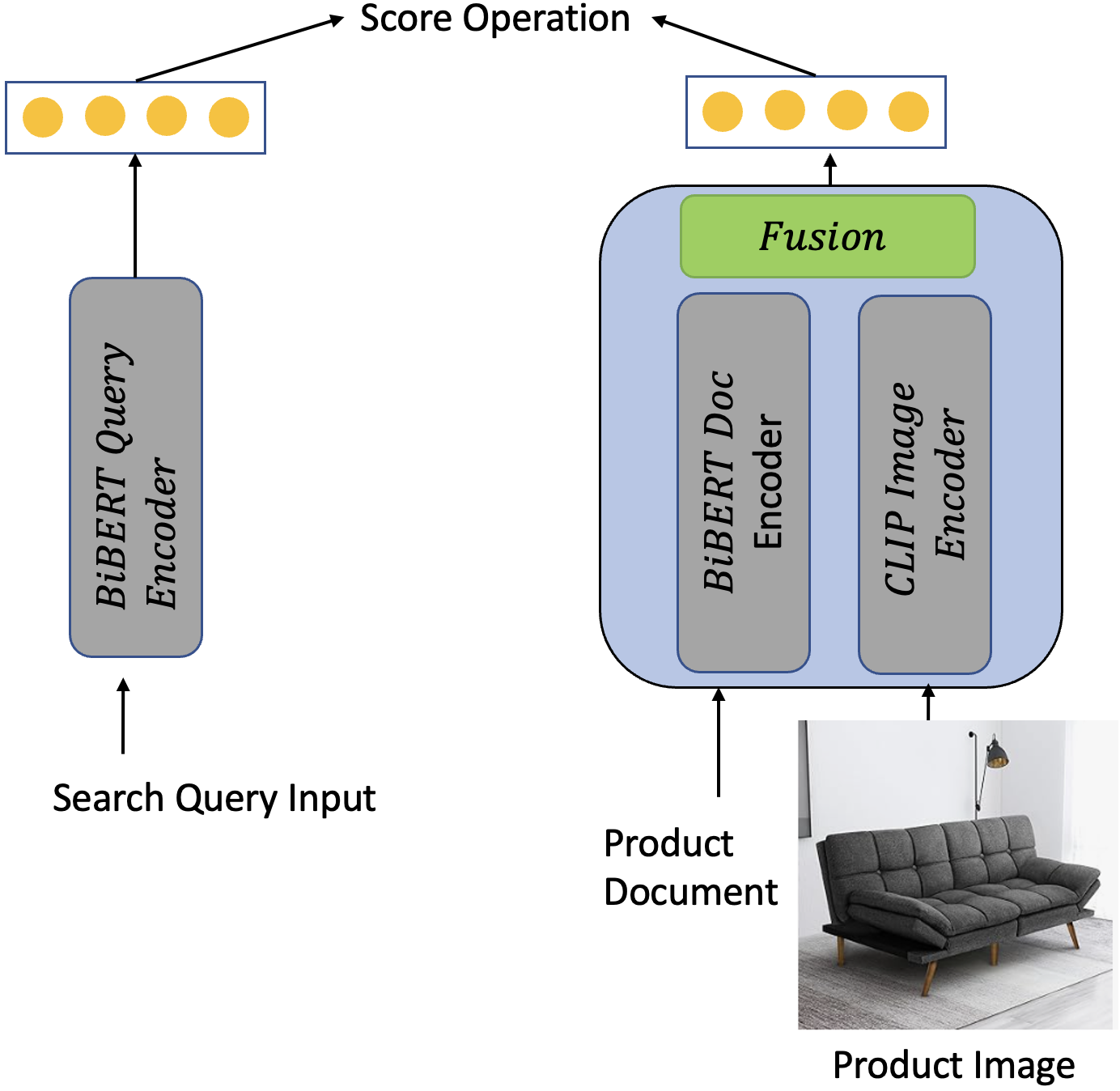} }}%
    \caption{Two multimodal model architectures for semantic retrival}
    \label{fig:multimodal}
\end{figure*}

\subsection{4-tower Multi-modality Model (4tMM)}

4tMM employs (refer to Figure~\ref{fig:multimodal}(a)) the baseline encoders of BiBERT model and CLIP model as the model backbone. The embeddings from different encoders are fused by a fusion module to form the final embedding representation of query and document. To be specific, denote the query and document encoders by $\phi_q^{BiBERT}$ and $\phi_d^{BiBERT}$, respectively. Similarly, let us denote the text and image encoder of CLIP model by $\phi_t^{CLIP}$ and $\phi_i^{CLIP}$, respectively. Then, with slightly abuse of notation, in the 4eMM architecture, we have the embedding representation of a query and a product as 
\begin{align}
    \bm{q} &= f(\phi_q^{BiBERT}(\bm{x}_{query}), \phi_t^{CLIP}(\bm{x}_{query})), \nonumber\\
    \bm{d} &= f(\phi_d^{BiBERT}(\bm{x}_{text}), \phi_i^{CLIP}(\bm{x}_{image})),
\end{align}
where $f(\cdot, \cdot)$ is a fusion mapping module to fuse the features from BiBERT and CLIP encoders into the final representation for search query (or similarly, for product). $\bm{x}_{query}$ denotes the search query input, $\bm{x}_{text}$ denotes the product item text input and $\bm{x}_{image}$ denotes the product image input. Note that the fusion mapping for query representation can be different from that of product representation.

We start with direct concatenation of fusion module,
\begin{align}
     f_{cat}(\bm{v}_1, \bm{v}_2) &= \bm{v}_1 \oplus \bm{v}_2
\end{align}
with $\oplus$ denoting concatenation for any given vector $\bm{v}_1 \in \mathbb{R}^{m}$ and $\bm{v}_2\in \mathbb{R}^{n}$. Apart from the direct concatenation, we also consider a $\alpha$-weighted concatenation as 
\begin{align}
     f_{\alpha-cat}(\bm{v}_1, \bm{v}_2) &= (\alpha \cdot \bm{v}_1) \oplus ((1-\alpha) \cdot \bm{v}_2), \hspace{1cm} \alpha \in (0,1).
\end{align}

To generalize the fusion module better, we also implement the multiple layer perception (MLP) fusion, i.e. $f_{MLP}(\cdot, \cdot) = MLP(\cdot, \cdot)$. In all variants of the 4-tower multimodal model, the fusion strategy  $f$ is symmetric implemented on both query and document representation side. The symmetric implementation here refers to the MLP architecture being the same on both side but weight of MLP module is not shared, i.e. when using MLP-fusion $f_{MLP}(\cdot, \cdot) = MLP(\cdot, \cdot)$, the query side MLP is independent from the product side MLP.

\subsection{3-tower based Multimodal Model (3tMM)}
\label{sec:semm}
Apart from the benchmarking of 4tMM with four towers of encoders, we are also interested in the training of a 3-tower multimodal model (refer to Figure~\ref{fig:multimodal}(b)) to understand how large the gap could be if we get ride of the text encoder tower of CLIP model. This also helps us understand how large the vector space misalignment is and if it can be closed by model fine-tuning.
In this case, with trivial notation abuse, our 3tMM becomes as

\begin{align}
    \bm{q} &= \phi_q^{BiBERT}(\bm{x}_{query}), \nonumber\\
    \bm{d} &= f_{MLP}(\phi_d^{BiBERT}(\bm{x}_{text}), \phi_i^{CLIP}(\bm{x}_{image})),
\end{align}
where the $f_{MLP}$ is functioning as to aligning the embedding of image to the BiBERT vector space, and as the same time to fusion the features from different modalities. In training 3tMM, we would benchmark both training MLP fusion only and jointly training of MLP, $\phi_q$ and $\phi_d$.

\begin{table*}
  \caption{Statics of Training and Evaluation Datasets}
  \label{data_statics}
  \centering
  \begin{tabular}{llll}
    \toprule
    Datasets     & \# query-product pairs     & \# distinct queries  & \# distinct products \\
    \midrule
    Training Data & 5,400,836  & 988,090   & 581,158     \\
    Evaluation Data     & 150,179      & 38,268  & 3,384,067 \\
    \bottomrule
  \end{tabular}
 
\end{table*}
\begin{table*}
 \caption{Benchmarks of Models (metrics reported in percentage)}
  \label{tab:model_statics}

  \centering
  \adjustbox{max width=\textwidth}{
  \begin{tabular}{lllllllll}
    \toprule
    Model     & Tune BiBERT     &  Train MLP  & Seen Negative & Recall@100 & Exact & Substitute & Complement & Irrelevant \\
    \midrule
    BiBERT  &&& & 78.1  &  52.7  & 30.3 & 2.4 & 13.6     \\
    CLIP  &&& & 46 & 45.4 & 26.5 & 2.7 & 25.4 \\
    \midrule
    4tMM cat &&&& \textbf{78.6} & 52.5 & 31.1 & 2.4 & 14  \\
    4tMM $\alpha$-cat &&&&  78.5 & 51.9 & \textbf{31.2} & 2.4 & 14.5 \\
    4tMM &  &\checkmark&& 75.5 & 51.3 & 31.1 & \textbf{2.8} & 14.8 \\
    4tMM & \checkmark & \checkmark && 73.3 & \textbf{54} & 26.8 & 2.7 & \textbf{11.9} \\
    4tMM & \checkmark & \checkmark &\checkmark& 77.8 & 50.6 & 30.5 & 2.3 & 13.6 \\
    4tMM &  & \checkmark &\checkmark& 76.4 & 51.5 & 30.2 & 2.3 & 13.3 \\
    \midrule
    3tMM & \checkmark & \checkmark & & 73.1 & 53.8 & 26.8 & 2.7 & 12.1 \\ 
    3tMM & \checkmark & \checkmark & \checkmark& 74.9 & 50 & 30.4 & 2.6 & 14 \\ 
    3tMM & & \checkmark & \checkmark& 76.8 & 52.1 & 29.5 & 2.4 & 13.2 \\ 
    \bottomrule
  \end{tabular}
}
\end{table*}

\section{Experiments}

\subsection{Training and Evaluation Dataset}
For experimentation, we sampled and aggregated query-product interaction data from search logs over a 12-month period for training dataset.
The relevance label is pruned from query-product engagement.
The evaluation dataset contains 38268 unique queries and 3384067 unique product items, which is produced by pruning one month of query-product interaction data where the time window has no overlapping with that of the training data collection time. The statics of the datasets are summarized in Table~\ref{data_statics}.

In the above explained training dataset, a model being trained could see positively relevant query-document pairs in a batch of training data. Since we employ the NT-Xent loss in Eq.~\ref{eq_infonce}, the in-batch negatives are good contrasting examples for the model to tell positive examples from negative examples. On top of this, we create a variant of the training dataset by appending three additional negative products for each query. We get the three additional negative examples from the product items mactched for a query but without purchase records. These additional negative examples per query enrich the original training dataset, allowing the model to see more negative example in each training batch. We refer to this data variant as Seen Negative. We would also conduct benchmarks on seen negative data to show its impact.  

\subsection{Experimental Setup}
Since the BiBERT model and CLIP model are pretrained separately on different data, the embeddings produced by one encoder is not aligned with the embeddings produced by another. Therefore, we conduct the alignment fine-tuning training for encoders of the BiBERT and CLIP models. We use the training data as stated in the previous section and the loss function in Equation~\ref{eq_infonce} to fine-tune the proposed 4tMM and 3tMM. We use the learning rate of $1e-4$ all fine-tuning jobs. According to the specific model tuning setup (as would be detailed in the following section), we enable the gradient flow to different parts of 4tMM and 3tMM models.

To conduct fair comparisons bettwen different benchmarks, we evaluate all models using the same  evaluation datset, i.e. using the same 38K queries for all inference evaluation and all the 3.38 million product items. Each evaluation consists of: i) generating semantic embeddings for all test queries and embeddings for all products; ii) using KNN (k-nearest neighbors) algorithm in FAISS library \footnote{https://github.com/facebookresearch/faiss} for top-100 relevant products retrieval; iii) computing the recall and relevance metrics. We train all our models on 8 NVIDIA V100 GPUs with total 128GB GPU memory. The project is done with pytorch lighgning framework and Adam optimizer.  

We define two evaluation metrics for our model performance comparisons. The recall is computed at the top-100 predictions per query: we compute the ratio of purchased products in the model top-100 predictions versus the actual number of purchases per query, which is then averaged over all queries in evaluation dataset. We also evaluate the benchmarks on the precision metric, relying on a product annotation model to get four types of relevance labels for each query-product pair: Exact, Substitute, Complement, Irrelevant. For each model we benchmark, we report the precision of each relevance label. Note in relevance judgement, there are query-product pairs that cannot be labelled. We omit the such pairs for the trivial number of them.

\subsection{Experimental Results}

The experimental results of benchmark comparisons are shown in Table~\ref{tab:model_statics}.
By checking out the performance of CLIP, it shows high exact and substitute matches by applying the CLIP alone, though its recall falls behind BiBERT. The performance of CLIP model alone is decent good since it is trained on public dataset \cite{fang2023data} and the model has not seen any product search dataset. It shows good generalization capacity of the encoder models of CLIP.

When we compare benchmarks of 4eMM models versus BiBERT, we see some recall@100 improvement from embedding concatenation (4eMM cat and 4eMM $\alpha$-cat) by roughly 40-50 basis points. Meanwhile, the precision metric is bit mixed: it shows gains in substitute but loss in exact and irrelevant. 

Then we study the impact of fine-tuning BiBERT or MLP fusion module. It shows that jointly fine-tuning BERT model of BiBERT and MLP fusion module gives the best precision: 54\% of exact matches and 11.9\% of irrelevant matches. But the recall of the model is degraded at the same time, reaching to 73.3\%. Additionally, benchmark of fine-tuning MLP fusion module alone underperforms BiBERT model mostly except for the precision at complement matches.  When looking at the impact of seen negative to 4eMM, it seems to be more helpful on recall than relevance precision.

As explained in Section~\ref{sec:semm}, we also studied if the fusion MLP module can help to align the embedding space between CLIP image encoder and the text encoder of BiBERT model. Thus our benchmarks also include the 3eMM performance. Overall the performance of 3eMM variants are alike the performance of 4eMM variants: i) jointly fine-tuning of BiBERT and MLP module helps to improve relevance precision a lot (improving exact matches at the same time of reducing the irrelevant matches) but gets worse recall; ii) seen negative examples in model training are helpful only to recall but not precision. 

\subsection{Further Analysis: Net Purchase Recall and Relevance}
\begin{table*}
  \caption{Analysis of Product Match Overlapping between Models }
  \label{tab:overlap}
  \centering  
  \begin{tabular}{lllllll}
    \toprule
    Model     &\begin{tabular}{@{}c@{}} BiBEERT vs  \\ CLIP \end{tabular}     
    & \begin{tabular}{@{}c@{}} BiBERT vs  \\ 4tMM cat \end{tabular} 
    & \begin{tabular}{@{}c@{}} CLIP vs  \\ 4tMM cat \end{tabular} 
    & \begin{tabular}{@{}c@{}} BiBERT vs  \\ 4tMM MLP tuned \end{tabular} \\
    \midrule
    Mean & 15 & 39 & 32 & 37 \\
    Standard deviation & 15 & 21  & 20 & 20 \\ 
    \bottomrule
  \end{tabular}
\end{table*}
\begin{table*}[ht]
  \caption{Analysis of Exclusive Matches}
  \label{tab:analysis}
  
  \centering
  \adjustbox{max width=\textwidth}{
  \begin{tabular}{lllllll}
    \toprule
    Model     &\begin{tabular}{@{}c@{}}\# Net Prediction  \\ (mean, standard deviation) \end{tabular}     &  \begin{tabular}{@{}c@{}} Net \\ Recall@100  \end{tabular}  & \begin{tabular}{@{}c@{}} Net \\ Exact \end{tabular}  & \begin{tabular}{@{}c@{}} Net \\ Substitute \end{tabular} & \begin{tabular}{@{}c@{}} Net \\ Complement \end{tabular} & \begin{tabular}{@{}c@{}} Net \\ Irrelevant \end{tabular} \\
    \midrule
    CLIP & (84, 15) & 10.1 & 39.1 & 31.5 & 3 & 31.5 \\
    4eMM cat & (60, 21) & 20.9 & 52.4 & 30.8 & 2.2 & 14.5  \\
    4eMM train MLP & (62, 20) & 17.7 & 50.1 & 31.1 & 3.1 & 15.8 \\
    4eMM BiBERT-MLP joinly trained & {(60, 23)} & {56.2} & {57.6} & {28.7} & {2.2} & {11.4} \\
    3eMM BiBERT-MLP joinly trained & {(59, 23)} & {56.2} & {57.6} & {28.7} & {2.2} & {11.5} \\
    3eMM train MLP $+$ Seen Negative & {(55, 24)} & {57.9} & {53.7} & {31.2} & {2.2} & {13.1} \\
    \bottomrule
  \end{tabular}
  }

\end{table*}

To understand if the developed models provide new information or not upon the baseline BiBERT, 
as shown in Table~\ref{tab:overlap}, the overlapping predictions are low when comparing against BiBERT predictions. The overlapping matches are lower for CLIP than 4tMM variants in contrast to BiBERT. 
We conduct further analysis of studying the statics, and metrics of “newly” predicted documents that are not in the matches set of BiBERT model. We define the "Net" metrics to measure the exclusive matches by a candidate model in contrast to baseline BiBERT model.
In order to measure the exclusive predicted products “quality”, we define the Net Purchase Recall and Relevance as follows: for a candidate model, i) we get the its KNN predictions, and compare the products predicted by the candidate model against products predicted by the baseline BiBERT model; ii) we exclude the baseline BiBERT model prediction product items from the candidate model’s predictions, namely the Net Predictions; iii) we evaluate the Net Predictions following ways:
\begin{itemize}
    \item Statics of counts of such Net Predictions, i.e. means and standard deviations.
    \item Net Purchase Recall: the Purchase Recall metric, computed against the ground-truth purchases
    \item The Net Relevance, evaluated by relevance annotation model Sentinel.
    
\end{itemize}


As shown in Table~\ref{tab:analysis}, the net predictions are pretty high (60 to 80 exclusive matches per query). Additionally, the exclusive matches still accounts for 10\% to 20\% recall when comparing the to ground-truth matched product documents. Note that precision of 4eMM variants is still very high, i.e. about 50\% exact matches and 30\% substitute matches. The results indicate that multimodal models are able to provide high-quality matches of products exclusively.

\section{Conclusion}

We developed multiple multimodal models to understand how product images make a difference to our product retrieval problem in e-commerce. After training a baseline of BiBERT for text-only retrieval, we start with benchmark of a CLIP model alone on product retrieval and it shows high relevance, i.e. 71.9\% mean Exact+Substitute but low purchase recall (46\%) in contrast to BiBERT. 
We then developed the variants of 4-tower and 3-tower multimodal models for product retrieval problem. We validated that a multimodal model can improve upon text-only BiBERT model on metrics of recall or relevance precision, depending on how we align the language and visual representation. Overall, our multimodal models show larger potential on improving relevance accuracy (e.g. higher exact relevant and lower irrelevant products measured in retrieval) than purchase prediction. A 4-tower model has better language-visual alignment than a 3-tower model and alignment gap can be relieved in model further training.
We additionally study the exclusive product matchset of our different benchmarks to gain further insights.
We compute the recall and relevance on the exclusive match set (net recall and relevance). We observe a high proportion of exacts and substitutes of our multimodal models and low proportion of irrelevants. The net recall is relatively low for the concatenation and MLP variants of 4tMM models but remains highly over 50\% for the joint training cases of 4tMM, 3tMM and adding better negative examples for 3tMM.  



\begin{acks}

We would like to thank to our colleagues who helped and supported this project: David Saile, Ronald Denaux, Paula Batlle, Virginia Fernandez Arguedas,  Biswadeep Das Roy, Mutasem Al-Darabsah, Suraj Nair, Choon Hui Teo, Subhendu Aich, Arnab Dhua, Jinyu Yang, Son Tran and Marcello Federico. 

\end{acks}

\medskip
\bibliographystyle{plain}
\bibliography{references}

\begin{thebibliography}{10}

\bibitem{alayrac2022flamingo}
Jean-Baptiste Alayrac, Jeff Donahue, Pauline Luc, Antoine Miech, Iain Barr,
  Yana Hasson, Karel Lenc, Arthur Mensch, Katherine Millican, Malcolm Reynolds,
  et~al.
\newblock Flamingo: a visual language model for few-shot learning.
\newblock {\em Advances in neural information processing systems},
  35:23716--23736, 2022.

\bibitem{chang2021extreme}
Wei-Cheng Chang, Daniel Jiang, Hsiang-Fu Yu, Choon~Hui Teo, Jiong Zhang, Kai
  Zhong, Kedarnath Kolluri, Qie Hu, Nikhil Shandilya, Vyacheslav Ievgrafov,
  et~al.
\newblock Extreme multi-label learning for semantic matching in product search.
\newblock In {\em Proceedings of the 27th ACM SIGKDD conference on knowledge
  discovery \& data mining}, pages 2643--2651, 2021.

\bibitem{chen2020simple}
Ting Chen, Simon Kornblith, Mohammad Norouzi, and Geoffrey Hinton.
\newblock A simple framework for contrastive learning of visual
  representations.
\newblock In {\em International conference on machine learning}, pages
  1597--1607. PMLR, 2020.

\bibitem{dosovitskiy2020image}
Alexey Dosovitskiy.
\newblock An image is worth 16x16 words: Transformers for image recognition at
  scale.
\newblock {\em arXiv preprint arXiv:2010.11929}, 2020.

\bibitem{fang2023data}
Alex Fang, Albin~Madappally Jose, Amit Jain, Ludwig Schmidt, Alexander Toshev,
  and Vaishaal Shankar.
\newblock Data filtering networks.
\newblock {\em arXiv preprint arXiv:2309.17425}, 2023.

\bibitem{huang2013learning}
Po-Sen Huang, Xiaodong He, Jianfeng Gao, Li~Deng, Alex Acero, and Larry Heck.
\newblock Learning deep structured semantic models for web search using
  clickthrough data.
\newblock In {\em Proceedings of the 22nd ACM international conference on
  Information \& Knowledge Management}, pages 2333--2338, 2013.

\bibitem{kenton2019bert}
Jacob Devlin Ming-Wei~Chang Kenton and Lee~Kristina Toutanova.
\newblock Bert: Pre-training of deep bidirectional transformers for language
  understanding.
\newblock In {\em Proceedings of naacL-HLT}, volume~1, page~2, 2019.

\bibitem{khattab2021relevance}
Omar Khattab, Christopher Potts, and Matei Zaharia.
\newblock Relevance-guided supervision for openqa with colbert.
\newblock {\em Transactions of the association for computational linguistics},
  9:929--944, 2021.

\bibitem{kim2021vilt}
Wonjae Kim, Bokyung Son, and Ildoo Kim.
\newblock Vilt: Vision-and-language transformer without convolution or region
  supervision.
\newblock In {\em International conference on machine learning}, pages
  5583--5594. PMLR, 2021.

\bibitem{lakshman2021embracing}
Vihan Lakshman, Choon~Hui Teo, Xiaowen Chu, Priyanka Nigam, Abhinandan Patni,
  Pooja Maknikar, and SVN Vishwanathan.
\newblock Embracing structure in data for billion-scale semantic product
  search.
\newblock {\em arXiv preprint arXiv:2110.06125}, 2021.

\bibitem{pmlr-v162-li22n}
Junnan Li, Dongxu Li, Caiming Xiong, and Steven Hoi.
\newblock {BLIP}: Bootstrapping language-image pre-training for unified
  vision-language understanding and generation.
\newblock In Kamalika Chaudhuri, Stefanie Jegelka, Le~Song, Csaba Szepesvari,
  Gang Niu, and Sivan Sabato, editors, {\em Proceedings of the 39th
  International Conference on Machine Learning}, volume 162 of {\em Proceedings
  of Machine Learning Research}, pages 12888--12900. PMLR, 17--23 Jul 2022.

\bibitem{li2021embedding}
Sen Li, Fuyu Lv, Taiwei Jin, Guli Lin, Keping Yang, Xiaoyi Zeng, Xiao-Ming Wu,
  and Qianli Ma.
\newblock Embedding-based product retrieval in taobao search.
\newblock In {\em Proceedings of the 27th ACM SIGKDD Conference on Knowledge
  Discovery \& Data Mining}, pages 3181--3189, 2021.

\bibitem{lin2021deep}
Qinghong Lin, Xiaojun Chen, Qin Zhang, Shangxuan Tian, and Yudong Chen.
\newblock Deep self-adaptive hashing for image retrieval.
\newblock In {\em Proceedings of the 30th ACM International Conference on
  Information \& Knowledge Management}, pages 1028--1037, 2021.

\bibitem{christof2020picture}
Christof Naumzik and Stefan Feuerriegel.
\newblock One picture is worth a thousand words? the pricing power of images in
  e-commerce.
\newblock In {\em Proceedings of The Web Conference 2020}, WWW '20, page
  3119–3125, New York, NY, USA, 2020. Association for Computing Machinery.

\bibitem{nie2020dc}
Ping Nie, Yuyu Zhang, Xiubo Geng, Arun Ramamurthy, Le~Song, and Daxin Jiang.
\newblock Dc-bert: Decoupling question and document for efficient contextual
  encoding.
\newblock In {\em Proceedings of the 43rd international ACM SIGIR conference on
  research and development in information retrieval}, pages 1829--1832, 2020.

\bibitem{radford2021learning}
Alec Radford, Jong~Wook Kim, Chris Hallacy, Aditya Ramesh, Gabriel Goh,
  Sandhini Agarwal, Girish Sastry, Amanda Askell, Pamela Mishkin, Jack Clark,
  et~al.
\newblock Learning transferable visual models from natural language
  supervision.
\newblock In {\em International conference on machine learning}, pages
  8748--8763. PMLR, 2021.

\bibitem{schutze2008introduction}
Hinrich Sch{\"u}tze, Christopher~D Manning, and Prabhakar Raghavan.
\newblock {\em Introduction to information retrieval}, volume~39.
\newblock Cambridge University Press Cambridge, 2008.

\bibitem{zhan2020repbert}
Jingtao Zhan, Jiaxin Mao, Yiqun Liu, Min Zhang, and Shaoping Ma.
\newblock Repbert: Contextualized text embeddings for first-stage retrieval.
\newblock {\em arXiv preprint arXiv:2006.15498}, 2020.

\bibitem{zhang2021binary}
Wanqian Zhang, Dayan Wu, Yu~Zhou, Bo~Li, Weiping Wang, and Dan Meng.
\newblock Binary neural network hashing for image retrieval.
\newblock In {\em Proceedings of the 44th international ACM SIGIR conference on
  research and development in information retrieval}, pages 1318--1327, 2021.

\bibitem{zhu2024bringing}
Xinliang Zhu, Sheng-Wei Huang, Han Ding, Jinyu Yang, Kelvin Chen, Tao Zhou, Tal
  Neiman, Ouye Xie, Son Tran, Benjamin Yao, et~al.
\newblock Bringing multimodality to amazon visual search system.
\newblock In {\em Proceedings of the 30th ACM SIGKDD Conference on Knowledge
  Discovery and Data Mining}, pages 6390--6399, 2024.

\bibitem{zhu2022uni}
Xizhou Zhu, Jinguo Zhu, Hao Li, Xiaoshi Wu, Hongsheng Li, Xiaohua Wang, and
  Jifeng Dai.
\newblock Uni-perceiver: Pre-training unified architecture for generic
  perception for zero-shot and few-shot tasks.
\newblock In {\em Proceedings of the IEEE/CVF Conference on Computer Vision and
  Pattern Recognition}, pages 16804--16815, 2022.

\end{thebibliography}

\appendix

\end{document}